\documentstyle[preprint,aps]{revtex}
\begin{document}
\tightenlines
\draft

\preprint{
\parbox{4cm}{
\baselineskip=12pt
TMUP-HEL-0014\\
UT-902\\ 
KEK-TH-703\\ 
July, 2000\\
\hspace*{1cm}
}}

\title{R-mediation of Dynamical Supersymmetry Breaking}
\author{ Noriaki Kitazawa $^a$ 
  \thanks{e-mail: kitazawa@phys.metro-u.ac.jp}, 
         Nobuhito Maru $^b$ 
  \thanks{e-mail: maru@hep-th.phys.s.u-tokyo.ac.jp, JSPS Research Fellow}
     and Nobuchika Okada $^c$ 
  \thanks{e-mail: okadan@camry.kek.jp, JSPS Research Fellow}}  
\address{$^a$Department of Physics, Tokyo Metropolitan University,
         Hachioji, Tokyo 192-0397, Japan}
\address{$^b$Department of Physics, University of Tokyo,
         Tokyo 113-0033, Japan}
\address{$^c$Theory Group, KEK, Tsukuba, Ibaraki 305-0801, Japan}
%\date{\today}
\maketitle

\begin{abstract}
We propose a simple scenario
 of the dynamical supersymmetry breaking
 in four dimensional supergravity theories.
The supersymmetry breaking sector
 is assumed to be completely separated
 as a sequestered sector from the visible sector,
 except for the communication
 by the gravity and U(1)$_R$ gauge interactions,
 and the supersymmetry breaking is mediated
 by the superconformal anomaly and U(1)$_R$ gauge interaction.
Supersymmetry is dynamically broken
 by the interplay between the non-perturbative effect
 of the gauge interaction
 and Fayet-Iliopoulos D-term of U(1)$_R$
 which necessarily exists in supergravity theories
 with gauged U(1)$_R$ symmetry.
We construct an explicit model
 which gives phenomenologically acceptable mass spectrum
 of superpartners with vanishing (or very small) cosmological constant.
\end{abstract}
%\pacs{}
\newpage

\section{Introduction}
\label{sec:intro}

Low energy supersymmetry may play an important role
 in solving many problems of the particle physics.
If it is the case,
 supersymmetry must be spontaneously broken,
 and all superpartners must have appropriate masses,
 since their effect is not observed yet.
Therefore, 
 finding a simple mechanism of the supersymmetry breaking
 and its mediation without any phenomenological problems
 is an important task.
If we believe low energy supersymmetry,
 it is natural to consider the supergravity framework.

The simplest scenario of the supersymmetry breaking
 and its mediation in supergravity theories
 is the gravity mediation with Polonyi potential
 in the hidden sector\cite{polonyi},
 but supersymmetry is not dynamically broken in this scenario.
Moreover,
 it is well known that the gravity mediation
 has a phenomenological problem:
 the degeneracy of squark masses at Planck scale
 is distorted by the quantum effect at low energies,
 which causes supersymmetric flavor problem.
There is another conceptual problem about the gravity mediation
 as pointed out in Ref.\cite{randall-sundrum1}:
 it is not the mediation by the gravity,
 but the mediation by the higher dimensional contact interactions
 introduced by hand.
Although it is possible that
 the superspace density, which defines the supergravity Lagrangian,
 contains infinite number of higher dimensional contact interactions
 so that the K\"ahler potential has a simple canonical form,
 the origin of these interactions is mysterious.

There is another possibility that 
 the visible sector and hidden sector are completely separated,
 namely, no contact interaction among them in the superspace density.
This situation would be naturally realized,
 if two sectors are confined in the different branes 
 separated in the direction of extra dimensions
 (now the hidden sector should be called as
 the sequestered sector\cite{randall-sundrum1}).
In this case
 the supersymmetry breaking at the sequestered sector
 is transmitted to the visible sector
 only through the superconformal anomaly
 \cite{randall-sundrum1,GLMR,BMP}.
In this anomaly mediation
 the masses of squarks highly degenerate at low energies
 and there is no supersymmetric flavor problem,
 but sleptons have negative masses ($m_{\mbox{\rm slepton}}^2<0$).
There are many attempts to solve this problem
 \cite{randall-sundrum1,pomarol-rattazzi,CLMP,KSS},
 and we usually need some additional fields
 which communicate two sectors.
In this paper
 we introduce this additional communication
 by gauging U(1)$_R$ symmetry in four dimensional supergravity theories
 \cite{chamseddine-dreiner,CFM,JJ}.
Since the charge of U(1)$_R$ symmetry
 does not commute with supercharges,
 it is natural to consider that
 the U(1)$_R$ gauge boson propagates in whole space-time
 including extra dimensions,
 and communicates two sectors.

It is also interesting to note that
 Fayet-Iliopoulos term for U(1)$_R$ must exist
 due to the symmetry of supergravity,
 and this term can act an important role
 in the supersymmetry breaking.
In fact
 it has been shown that the supersymmetry can be dynamically broken
 by the interplay between this Fayet-Iliopoulos term
 and the non-perturbative effect of a gauge interaction
 \cite{KMO}.
Since the auxiliary field of U(1)$_R$ gauge multiplet
 has vacuum expectation value,
 both squarks and sleptons can have positive masses
 of the order of the gravitino mass
 in an appropriate R-charge assignment,
 and the problem of the anomaly mediation can be avoided.

This paper is organized as follows.
In the next section
 we give a general argument
 on the supergravity Lagrangian with U(1)$_R$ gauge symmetry.
We give
 a general formula for the chirality-conserving scalar mass
 in the presence of U(1)$_R$ gauge symmetry,
 which is an extension of the formula given in Ref.\cite{IKYY}.
An explicit model is constructed in section 3,
 and the analysis of the dynamics and mass spectrum
 is given in section 4.
Section 5 contains our conclusions.

\section{Supergravity with U(1)$_R$ gauge symmetry}
\label{sec:sugra}

In the superconformal framework\cite{kaku-townsend,kugo-uehara,FGKP}
 the general supergravity Lagrangian
 with U(1)$_R$ gauge symmetry is given by
\begin{eqnarray}
 {\cal L}
 = &-& {1 \over 2}
     \left[
      {\bar \Sigma}_c e^{-2 g_R V_R} \Sigma_c
       \Phi (S_I, {\bar S}^I e^{2 Q_I g_R V_R} e^{2 g_G V_G})
     \right]_D
\nonumber\\
   &+& \left[ W(S_I) \Sigma_c^3 \right]_F
   - {1 \over 4} \left[ f_R(S_I) W_R W_R \right]_F
   - {1 \over 4} \left[ f_{ab}(S_I) W_G^a W_G^b \right]_F,
\end{eqnarray}
 where we use the notation in Ref.\cite{kugo-uehara}.
Here,
 $S_I$ are matter chiral multiplets
 with flavor index $I$ and U(1)$_R$ charge $Q_I$,
 $V_R$ and $V_G$ ($W_R$ and $W_G$) are vector (chiral) multiplets
 corresponding to the gauge group of U(1)$_R$ and $G$, respectively.
The multiplet $\Sigma_c$ is the compensating multiplet,
 whose component should be appropriately fixed to obtain
 Poincar\'e supergravity.
The functions $\Phi$ and $W$ are superspace densities
 in which interactions are described by the products of multiplets.
Following the arguments in the previous section,
 we assume that there is no interaction
 between the visible sector fields $S_i$ and $V_{Gv}$
 and the hidden (sequestered) sector fields $S_\alpha$ and $V_{Gh}$
 in these superspace densities, namely,
\begin{equation}
 \Phi (S_I, {\bar S}_I e^{2 Q_I g_R V_R} e^{2 g_G V_G})
 = \Phi_v
    (S_i, {\bar S}^i e^{2 Q_i g_R V_R} e^{2 g_{Gv} V_{Gv}})
 + \Phi_h
    (S_\alpha,
    {\bar S}^\alpha e^{2 Q_\alpha g_R V_R} e^{2 g_{Gh} V_{Gh}}),
\label{D-density}
\end{equation}
\begin{equation}
 W(S_I) = W_v (S_i) + W_h (S_\alpha),
\label{F-density}
\end{equation}
 where indices $i$ and $\alpha$
 denote the flavors in the visible and hidden sectors, respectively,
 and $Gv$ and $Gh$ are gauge groups in each sector.
The gauge kinetic function $f_{ab}(S_I)$
 should also be restricted as follows.
\begin{equation}
 \left[ f_{ab}(S_I) W_G^a W_G^b \right]_F
 \longrightarrow
 \left[ f_{ab}^{Gv} (S_i) W_{Gv}^a W_{Gv}^b \right]_F
 + \left[ f_{ab}^{Gh} (S_\alpha) W_{Gh}^a W_{Gh}^b \right]_F.
\end{equation}
In the following
 we assume $f_R(S_I)=1$ and
 $f_{ab}^{Gv}=f_{ab}^{Gh}=\delta_{ab}$, for simplicity.

Note that
 the compensating multiplet $\Sigma_c$ must have R-charge,
 since the superpotential $W$ has R-charge.
Therefore, 
 the usual gauge choice to give Poincar\'e supergravity:
\begin{eqnarray}
 z_c = \sqrt{3}, \ \
 \chi_{Rc} = 0, \ \
 b_\mu = 0
\end{eqnarray}
 does not preserve U(1)$_R$ symmetry,
 where $z_c$ and $\chi_{Rc}$ are scalar and spinor components
 of the compensating multiplet $\Sigma_c$
 and $b_\mu$ is one of the gauge fields
 of the superconformal gauge group.
We have to rescale the compensating multiplet
 to obtain the R-symmetric Poincar\'e supergravity:
\begin{equation}
 S_0 \equiv \Sigma_c \left( W(S_I) \right)^{1/3}.
\end{equation}
The Lagrangian becomes
\begin{eqnarray}
 {\cal L}
 = &-& {1 \over 2}
     \left[
      {\bar S}_0 S_0
      {\tilde \Phi} (S_I, {\bar S}^I e^{2 Q_I g_R V_R} e^{2 g_G V_G})
     \right]_D
   + \left[ S_0^3 \right]_F
\nonumber\\
   &-& {1 \over 4} \left[ W_R W_R \right]_F
   - {1 \over 4} \left[ W_{Gv} W_{Gv} \right]_F
   - {1 \over 4} \left[ W_{Gh} W_{Gh} \right]_F,
\end{eqnarray}
 where
\begin{equation}
 {\tilde \Phi} (S_I, {\bar S}^I e^{2 Q_I g_R V_R} e^{2 g_G V_G}) \equiv
 {{\Phi (S_I, {\bar S}^I e^{2 Q_I g_R V_R} e^{2 g_G V_G})}
  \over
  {\left(
   {\bar W} ({\bar S}^I e^{2 Q_I g_R V_R} e^{2 g_G V_G}) W(S_I)
  \right)^{1/3}}}.
\label{Phi}
\end{equation}
The compensating multiplet $S_0$ is U(1)$_R$ singlet now.
It was shown in Ref.\cite{kugo-uehara}
 that the gauge fixing conditions of
\begin{equation}
 z_0 = \sqrt{3} {\tilde \Phi}^{-1/2} (z_I, z^*_I), \ \
 \chi_{R0} = - z_0 {\tilde \Phi}^{-1} {\tilde \Phi}^J \chi_{RJ}, \ \
 b_\mu = 0
\end{equation}
 directly give the standard form of the supergravity Lagrangian
 given in Ref.\cite{CFGP},
 where $z_0$ and $\chi_{R0}$ are scalar and spinor components
 of the compensating multiplet $S_0$,
 ${\tilde \Phi}^J
 \equiv \partial {\tilde \Phi} (z_I,z^{*I}) / \partial z_J$
 and $z_J$ is the scalar components of $S_J$.
After all, 
 the resultant Lagrangian in component fields
 has the standard form of Ref.\cite{CFGP}
 including covariant derivatives for U(1)$_R$ gauge symmetry.
The Lagrangian is determined by a function
\begin{equation}
 G(z_I,z^{*I}) \equiv -3 \ln {\tilde \Phi}(z_I,z^{*I})
 = -3 \ln \Phi(z_I,z^{*I}) + \ln | W(z_I) |^2,
\end{equation}
 where $\Phi$ and $W$ satisfy the conditions of
 Eqs.(\ref{D-density}) and (\ref{F-density}).
The difference of R-charges in covariant derivatives
 for each component field in a multiplet automatically appears
 due to the fact that $W$ has non-trivial R-charge
 (see Ref.\cite{chamseddine-dreiner}).

The potential for scalar fields is given as follows.
\begin{equation}
 V = V_F + V_D,
\end{equation}
 where F-term contribution is
\begin{equation}
 V_F = e^G \left( G_I (G^{-1})^I_J G^J - 3 \right)
\end{equation}
 and U(1)$_R$ D-term contribution is
\begin{equation}
 V_D = {{g_R^2} \over 2} \left( G^I Q_I z_I \right)^2.
\end{equation}
We take the reduced Planck scale as a unit of the mass scale.
The chirality-conserving scalar mass can be obtained
 by differentiating this potential by $z_i$ and $z^{*j}$
 and taking its vacuum expectation value.
In addition to the conditions of
 Eqs.(\ref{D-density}) and (\ref{F-density}),
 we introduce the conditions of
\begin{equation}
 \langle \Phi_{vi} \rangle, \
 \langle W_{vi} \rangle, \
 \langle z_i \rangle
 = 0 \ \ \mbox{\rm or} \ll 1.
\end{equation}
These conditions mean the assumption that
 the breaking scales of gauge symmetries in the visible sector
 should be much smaller than the reduced Planck scale.
We obtain,
\begin{equation}
 \langle V_F{}^i_j \rangle
 = m^{ik} \langle \left( G^{-1} \right)^l_k \rangle m_{lj}
 + {2 \over 3} \langle V_F \rangle \langle G^i_j \rangle,
\end{equation}
\begin{equation}
 \langle V_D{}^i_j \rangle
 = {2 \over 3} \langle V_D \rangle \langle G^i_j \rangle
 + g_R^2 \left( Q_i - {2 \over 3} \right)
   \langle D \rangle \langle G^i_j \rangle,
\end{equation}
 where $m^{ik}$ is the supersymmetric mass
 and $D \equiv G^I Q_I z_I$.
The superpotential $W$ has R-charge 2 in our convention.
Therefore,
 the chirality-conserving supersymmetry-breaking scalar mass
 is obtained as
\begin{equation}
 {\tilde m}^2_R{}^i_j
 = \left(
    {2 \over 3} \langle V \rangle
     + g_R^2 \left( Q_i - {2 \over 3} \right)
   \langle D \rangle
   \right) \langle G^i_j \rangle.
\label{r-mediation}
\end{equation}
The vacuum expectation value of the potential itself
 corresponds to the cosmological constant
 which should vanish in realistic models.
We see that there is no gravity mediation,
 but there is ``R-mediation''
 which is the tree-level contribution due to $\langle D \rangle \ne 0$.

\section{Constructing a model}
\label{sec:model}

We construct an explicit model
 to show that the scenario which is described in the first section
 is possible.
The particle contents of the model
 are summarized in Table \ref{contents}.
In the following
 we simply introduce the role of each field
 without mentioning the dynamics in detail.
The dynamics will be discussed in the next section.

As for the hidden sector,
 we take the same system which was introduced in Ref.\cite{KMO}.
It consists of
 two fields, $Q_1$ and $Q_2$,
 in the fundamental representation of SU(2)$_H$ gauge group
 and a Yukawa interaction with a SU(2)$_H$ singlet field $S$:
\begin{equation}
 W_h = \lambda S [Q_1 Q_2],
\end{equation}
 where square brackets denote the contraction of SU(2)$_H$ indices.
Supersymmetry is dynamically broken
 by the interplay between
 the non-perturbative effect of SU(2)$_H$ interaction
 and U(1)$_R$ Fayet-Iliopoulos term,
 if there is no other field with R-charge less than $2/3$.

The visible sector is based on
 the system of the minimal supersymmetric standard model.
At least the R-charges of leptons must be larger than $2/3$
 so that sleptons obtain positive masses
 through R-mediation of Eq.(\ref{r-mediation})
 (assuming $\langle D \rangle > 0$ in this section).
We simply assume that all the quarks and leptons have unit R-charge.
This means that the R-charges of Higgs fields
 must be zero (note that this is less than $2/3$),
 since we need the Yukawa couplings of
\begin{equation}
 W_v^Y = g_u H Q {\bar U} + g_d {\bar H} Q {\bar D}
       + g_\nu H L {\bar N} + g_e {\bar H} L {\bar E},
\label{higgs-yukawa}
\end{equation}
 where we suppress the generation indices, for simplicity.

To ensure the dynamical supersymmetry breaking
 we introduce other two Higgs fields, $H'$ and ${\bar H}'$,
 with the mass terms of
\begin{equation}
 W'_v = \mu_u H {\bar H}' + \mu_d {\bar H} H'. 
\label{higgs-mass}
\end{equation}
Although there are negative contributions
 to the masses of $H$ and ${\bar H}$ from Eq.(\ref{r-mediation}),
 these mass terms can make all masses of Higgs fields positive
 at the tree level.
Therefore,
 the electroweak symmetry must be broken radiatively\cite{radiative}.

At this stage,
 all the gauge anomalies are cancelled out,
 except for
 $\left( \mbox{\rm SU(3)}_c \right)^2 \mbox{\rm U(1)}_R$, 
 $\left( \mbox{\rm SU(2)}_L \right)^2 \mbox{\rm U(1)}_R$,
 $(\mbox{\rm U(1)}_R)^3$ and $\mbox{\rm U(1)}_R (\mbox{\rm gravity})^2$
 anomalies.
To cancel
 $\left( \mbox{\rm SU(3)}_c \right)^2 \mbox{\rm U(1)}_R$ and
 $\left( \mbox{\rm SU(2)}_L \right)^2 \mbox{\rm U(1)}_R$ anomalies,
 we further introduce additional fields, $\Omega_i$ and $\Sigma_i$,
 and Yukawa interactions with $X$.
\begin{equation}
 W_v^X
 = g X \left( \Omega_i \Omega_i + \Sigma_i \Sigma_i \right)
 + m X X'.
\label{superpotential-X}
\end{equation}
The field $X'$ and the mass term with $X$
 are required to have positive mass for $X$
 and to ensure the dynamical supersymmetry breaking,
 since $X$ has R-charge less than $2/3$.
The fields $\Omega_i$ and $\Sigma_i$ become heavy
 by the vacuum expectation value of $X$
 which is generated by the 1-loop effect of the Yukawa coupling
 in Eq.(\ref{superpotential-X}).
The remaining anomalies,
 $(\mbox{\rm U(1)}_R)^3$ and $\mbox{\rm U(1)}_R (\mbox{\rm gravity})^2$,
 can be cancelled out by introducing, for example,
 many fields of R-charge two with appropriate values of $q_1$ and $q_2$.
There may be much more sophisticated and convincing way
 to cancel these anomalies, but we leave it for further studies.

\section{Dynamics of the model}
\label{sec:dynamics}

Before discussing the dynamics of the model in detail,
 we have to make an assumption about the superspace density, $\Phi$.
We simply assume as
\begin{equation}
 \Phi (z_I, z^{*I}) = 1 - {1 \over 3} \sum_I z^{*I} z_I
\end{equation}
 respecting the condition of Eq.(\ref{D-density}),
 where $z_I$ are the scalar components
 of all the chiral multiplets in the model.
This gives canonical kinetic terms
 in the first order of the $1/M_P$ expansion,
 where $M_P=M_{\rm planck}/\sqrt{8\pi}$ is the reduced Planck scale.
In this case the scalar potential can be written as
\begin{equation}
 V = V_F + V_D
\end{equation}
 with
\begin{equation}
 V_F
 = {1 \over {\Phi^2}}
   \left\{
    W^*_I W^I
    - {1 \over 3} | z_I W^I |^2
    + \left( W^* W^I z_I + W W^*_I z^{*I} \right)
    - 3 | W |^2
   \right\},
\end{equation}
\begin{equation}
 V_D
 = {1 \over {\Phi^2}} {{g_R^2} \over 2}
   \left\{
    \left( Q_I - {2 \over 3} \right) z^{*I} z_I + 2
   \right\}^2,
\end{equation}
 where we neglect the D-term contributions
 form other gauge interactions.

First, we discuss the dynamics of the supersymmetry breaking.
The instanton effect of SU(2)$_H$ gauge interaction
 can be described as a dynamically generated superpotential
 \cite{ADS}.
The effective superpotential for the hidden sector is
\begin{equation}
 W_h^{\rm eff}
 = \lambda S [Q_1 Q_2] + {{\Lambda^5} \over {[Q_1 Q_2]}},
\end{equation}
 where $\Lambda$ is the scale of the dynamics of SU(2)$_H$.
If we assume that
 the vacuum expectation values of $Q_1$ and $Q_2$
 lie on the flat direction of SU(2)$_H$,
 we have
\begin{eqnarray}
 V_F
 &=& {1 \over {\Phi^2}}
   \left\{
    \left( \lambda v^2 \right)^2
    + 2 v^2 \left( \lambda s - {{\Lambda^5} \over {v^4}} \right)^2
    - {{25} \over 3}
      \left( {{\Lambda^5} \over {v^2}} \right)^2
    + \mbox{(visible sector)}
   \right\},
\\
 V_D
 &=& {{g_R^2} \over 2} D^2,
\end{eqnarray}
 where
\begin{equation}
 \Phi = 1 - {1 \over 3} s^2 - {2 \over 3} v^2
          - {1 \over 3}
            \left[ z^{*I} z_I \right]_{\rm visible},
\end{equation}
\begin{equation}
 D =
   {1 \over \Phi}
   \left\{ 
    \left( q_S - {2 \over 3} \right) s^2
    + \left( q_1 + q_2 - {4 \over 3} \right) v^2
    + 2
    + \left[
       \left( Q_I - {2 \over 3} \right) z^{*I} z_I
      \right]_{\rm visible}
   \right\},
\end{equation}
 $v$ describes the flat direction of SU(2)$_H$,
 $s$ and $q_S$ are the vacuum expectation value and R-charge of $S$
 ($q_S = 4$ and $q_1 + q_2 = -2$).
It can be shown that
 all the visible sector fields do not have vacuum expectation values
 at the tree level.
It is rather trivial for the fields with R-charge larger than $2/3$,
 but it is non-trivial for the fields $H$, $\bar H$ and $X$,
 since the vacuum expectation values of these fields
 negatively contribute to the vacuum energy in $V_D$.
These fields do not have vacuum expectation values at the tree level,
 if the mass parameters $\mu_u$, $\mu_d$ and $m$
 are appropriately large,
 as it will be explained at the end of this section.
Therefore, in the following
 we consider the stationary conditions for $v$ and $s$
 neglecting all the contributions from the visible sector.

The analysis is almost the same as in Ref.\cite{KMO}.
In case of $g_R^2 \gg \lambda \sim \Lambda^5$
 there should be a solution of stationary conditions
 so that $v \simeq \sqrt{3/5}$ and $s \ll 1$,
 which results almost vanishing $D$.
In this case the scalar potential approximately becomes
\begin{equation}
 V \simeq {1 \over {\Phi^2}} \left( {3 \over 5} \right)^2
   \left\{
    \lambda^2 - 3 \left( {5 \over 3} \right)^5 \Lambda^{10}
   \right\}.
\end{equation}
Therefore, we can expect that
 there is a solution of vanishing (or vary small) cosmological constant
 with $\lambda \sim \sqrt{5} (5/3)^2 \Lambda^5 \sim 6.2 \Lambda^5$.
Indeed,
 we can approximately obtain such a solution as
\begin{eqnarray}
 v &\simeq& \sqrt{{3 \over 5}} + {\sqrt{15} \over 6} s^2
   - {1 \over {g_R^2}}
     {{243 \lambda^2 + 6250 \Lambda^{10}}
      \over
      {900 \sqrt{15}}},
\\
 s &\simeq& {{675 \lambda \Lambda^5}
             \over
             {486 \lambda^2 + 6250 \Lambda^{10}}}
\end{eqnarray}
 with vanishing cosmological constant
 by tuning $\lambda \simeq 6.9 \Lambda^5$.
A complete numerical analysis gives a solution
\begin{equation}
 v \simeq \sqrt{3/5} + 0.012, \ \ s \simeq 0.14
\end{equation}
 with vanishing cosmological constant,
 where $g_R = 10^{-12}$,
 $\Lambda = 10^{-3}$ and $\lambda \simeq 6.9 \Lambda^5$.
At this vacuum the gravitino mass $m_{3/2}$ becomes
\begin{equation}
 m_{3/2} \equiv \langle e^{G/2} \rangle
 \simeq 5.0 \times {{\Lambda^5} \over {M_P^4}}.
\end{equation}
The contribution to the mass of the scalar field
 due to $\langle D \rangle \ne 0$  
 can also be obtained from Eq.(\ref{r-mediation}) as
\begin{equation}
 {\tilde m}_R^2 (Q)
 = g_R^2 \langle D \rangle \left( Q - {2 \over 3} \right)
 \simeq \left( 7.2 \times {{\Lambda^5} \over {M_P^4}} \right)^2
 \left( Q - {2 \over 3} \right),
\label{scalar-r}
\end{equation}
 where $Q$ is the R-charge of the scalar field.
We see that
 these supersymmetry-breaking masses
 are the same order of magnitude.
The phenomenologically acceptable values of these masses
 can be obtained by changing the value of $\Lambda$
 within the same order of magnitude.

We summarize the spectrum of the supersymmetry-breaking masses
 and other supersymmetry breaking terms in the visible sector.

Gauginos in the visible sector
 can have masses only through the anomaly mediation,
 since there should be no hidden (sequestered) sector field
 in the gauge kinetic function.
Therefore,
\begin{equation}
 m_{\lambda_i} \simeq {{\beta (g_i^2)} \over {2 g_i^2}} m_{3/2},
\label{gaugino-mass}
\end{equation}
 where $g_i$ and $\beta (g_i^2)$
 are the gauge coupling and its beta function
 of the gauge group $i$ in the visible sector, respectively
(see Ref.\cite{BMP} for more precise formula).
There are two contributions to the scalar mass:
\begin{equation}
 {\tilde m}^2
 \simeq
 - {1 \over 4} {{d \gamma} \over {d \ln \mu}} m_{3/2}^2
 + {\tilde m}_R^2 (Q),
\label{scalar-mass}
\end{equation}
 where $\mu$ is the renormalization scale,
 and $\gamma$ and $Q$ are the anomalous dimension and R-charge
 of the scalar field, respectively.
The first term is the contribution by the anomaly mediation
 and the second term is the contribution by R-mediation
 given by Eq.(\ref{scalar-r}). 
The second contribution always dominates the first contribution,
 since the second contribution is the tree-level one.
Therefore,
 the scalar filed with $Q > 2/3$ naturally has positive mass.
If we take $m_{3/2} \sim 10$TeV
 to have gaugino masses heavier than about $100$GeV,
 the scalar mass becomes of the order of $(10\mbox{\rm TeV})^2$.

Other supersymmetry breaking terms
 also emerge through the anomaly mediation.
The A-term emerges corresponding to each Yukawa coupling
 through the anomaly mediation:
\begin{equation}
 A_{\Phi_1 \Phi_2 \Phi_3}
 = - {1 \over 2}
     \left(
      \gamma_{\Phi_1} + \gamma_{\Phi_2} + \gamma_{\Phi_3}
     \right) m_{3/2},
\end{equation}
 where the Yukawa coupling of
 $W_{\rm Yukawa} = \lambda \Phi_1 \Phi_2 \Phi_3$
 is considered, and $\gamma$ denotes the anomalous dimension
 of each field.
The B-term emerges corresponding to each mass term at the tree level,
 since the mass term explicitly breaks the superconformal symmetry:
\begin{equation}
 B \simeq - m_{3/2}.
\end{equation}
Note that
 the order of the magnitude of $A$ is always smaller than that of $B$,
 since the B-term emerges at the tree level.

Next,
 we discuss the radiative electroweak symmetry breaking
 in our model.
When we neglect the hidden sector,
 Higgs fields do not have vacuum expectation values at the tree level,
 if the following conditions are satisfied.
\begin{eqnarray}
 \left( \mu_u^2 + {\tilde m}^2_R(q_H) \right)
 \left( \mu_u^2 + {\tilde m}^2_R(q_{H'}) \right) - \mu_u^2 B^2 > 0,
\\
 \left( \mu_d^2 + {\tilde m}^2_R(q_H) \right)
 \left( \mu_d^2 + {\tilde m}^2_R(q_{H'}) \right) - \mu_d^2 B^2 > 0,
\end{eqnarray}
 where $q_H$ is the R-charge of $H$ and ${\bar H}$
 and $q_{H'}$ is the R-charge of $H'$ and ${\bar H}'$.
This condition is satisfied,
 if $\mu_u$ and $\mu_d$ are slightly larger than $m_{3/2}$.
On the other hand,
 $\mu_u^2$ and $\mu_d^2$ must be larger than
 $|{\tilde m}^2_R(q_H)| \simeq {2 \over 3} m_{3/2}^2$
 so that our mechanism of the dynamical supersymmetry breaking works.
Therefore,
 it is natural to consider that the masses
 $\mu_u$ and $\mu_d$ are slightly larger than $m_{3/2}$
 and the electroweak symmetry breaking occurs radiatively
 at the 1-loop level
 through the large Yukawa coupling of the top quark
 and the supersymmetry-breaking mass of the scalar top.
We can analytically show that
 the radiative electroweak symmetry breaking
 at the weak scale is possible in this model.

Finally,
 we discuss the radiative mass generation for $\Omega_i$ and $\Sigma_i$
 which are introduced for the gauge anomaly cancellation.
These fields should become heavy so that
 running gauge couplings do not blow up
 before Planck scale. 
The point is whether $X$ has vacuum expectation value or not,
 since $\langle X \rangle \ne 0$ make these fields massive
 through the first term of Eq.(\ref{superpotential-X}).
When we neglect the hidden sector,
 the vacuum expectation value of $X$ vanishes at the tree level,
 if the following condition is satisfied.
\begin{equation}
 \left( m^2 + {\tilde m}^2_R (q_X) \right)
 \left( m^2 + {\tilde m}^2_R (q_{X'}) \right)
 - m^2 B^2 > 0,
\end{equation}
 where $q_X$ and $q_{X'}$ are the R-charges of $X$ and $X'$,
 respectively.
This condition is satisfied, if $m$ is larger than $m_{3/2}$.
On the other hand,
 $m^2$ must be larger than
 $|{\tilde m}^2_R(q_X)| \simeq {1 \over 6} m_{3/2}^2$
 so that our mechanism of the dynamical supersymmetry breaking works.
Therefore, if we naturally take $m^2 > m_{3/2}^2$,
 $X$ can not have vacuum expectation value at the tree level.
But, it can have vacuum expectation value radiatively
 at the 1-loop level by the large coupling $g$
 and the supersymmetry-breaking masses
 of the scalar components of $\Omega_i$ and $\Sigma_i$.
We can analytically show that
 the value of $g \langle X \rangle$ can be large enough
 ($g \langle X \rangle \simeq 10^{-2} M_P$, for example)
 with $m \sim m_{3/2}$, $g \sim 1$ and very small $g_R \sim 10^{-12}$.

\section{Conclusion}
\label{sec:conclusion}

We have proposed a sequestered sector scenario
 in which the supersymmetry breaking are mediated
 by the superconformal anomaly and U(1)$_R$ gauge interaction
 without gravity mediation at the tree level.
We constructed an explicit model
 in which supersymmetry is dynamically broken by the interplay
 between the non-perturbative effect of the gauge interaction
 and Fayet-Iliopoulos term of U(1)$_R$.
It was found that
 the problem of the tachyonic slepton in the anomaly mediation
 can be avoided in this scenario.
The spectrum of the supersymmetry breaking masses
 is very simple and there is no supersymmetric flavor problem.
We have also proposed a mechanism
 to radiatively generate the mass of the field
 which should not appear at low energies.

We mention a remarkable fact in this model:
 R-parity is not necessary.
In the minimal supersymmetric standard model
 R-parity is usually assumed to forbid interactions
 which violate baryon number symmetry.
In our model
 U(1)$_R$ gauge symmetry naturally
 act the same or rather stronger role.
It forbids in the superpotential
 not only renormalizable terms
 but also all the higher dimensional terms
 which violate baryon number symmetry.
This is a simple realization
 of the idea proposed in Ref.\cite{extra-U(1)}.

We briefly summarize phenomenological consequences of this model.
All the gauginos have the masses of the order of $100$GeV,
 and the lightest superparticle would be a neutralino (bino or zino).
Further understanding of the gaugino spectrum
 and the nature of the lightest superparticle
 requires more detailed analysis on the radiative correction
 to the spectrum as described in Ref.\cite{GLMR}.
All the scalar fermions have the masses of the order of $10$TeV.
Therefore,
 in near future collider experiments
 we could not discover scalar fermions, but gauginos.
The Higgs sector in this model is very different
 from the one in the minimal supersymmetric standard model,
 since it includes four Higgs doublets.
There would be three charged Higgs and seven neutral Higgs,
 and all of them would have the masses of the order of $10$TeV,
 except for one CP-even neutral Higgs
 which would have the mass of the order of the weak scale.
Therefore,
 we could see one Higgs boson in near future collider experiments,
 but it would be impossible to see other Higgs particles.

There is an important point
 which have to be investigated in future.
That is to derive the four dimensional effective theory
 from the fundamental theory in higher dimension.
For example,
 if we consider a five dimensional theory as a fundamental theory,
 we have to integrate out the degrees of freedom
 which can propagate in fifth dimension.
In our scenario
 such degrees of freedom are gravity and R gauge interaction.
Especially,
 it have to be investigated
 how an U(1) component of the larger R gauge symmetry in five dimensions
 is projected out to U(1)$_R$ gauge symmetry
 in four dimensional effective theories.
It is also to be investigated
 how completely two sectors are separated in the superspace densities
 in four dimensional effective theories.

Although we still do not have the comprehensive analysis
 about deriving four dimensional effective theories
 from higher dimensional gauged supergravity theories,
 it is possible to expect that
 the very small value of the U(1)$_R$ gauge coupling in this model
 could be naturally obtained in the large extra dimension scenario.
The result of Ref.\cite{MP} suggests that 
 the natural (dimensionful) value
 of the gauge coupling in the higher dimensional theory
 could naturally result very small (dimensionless) value
 of the gauge coupling in the four dimensional effective theory,
 if the extra dimensions have relatively large volume.
Other gauge and Yukawa couplings in this model
 are not suppressed by this mechanism,
 since the visible and sequestered sector
 are assumed to be confined in each 3-brane
 and only the supergravity multiplet and U(1)$_R$ gauge boson
 can propagate the bulk.
If the volume of the extra dimensions is very large,
 the U(1)$_R$ gauge boson and graviton would be observed
 in near future collider experiments.

Finally,
 we want to emphasize that
 the proposed scenario is very simple,
 and we can rather easily construct calculable models
 which have concrete predictions.
We believe that this direction is worth investigating further.

\acknowledgments

This work was supported in part 
 by the Grant-in-aid for Science and 
 Culture Research from the Ministry of Education, 
 Science and Culture of Japan (\#11740156, \#08557, \#2997). 
N.M. and N.O. are supported by the Japan Society for 
 the Promotion of Science for Young Scientists.

\begin{table}
\begin{tabular}{cccccc}
            & SU(3)$_c$ & SU(2)$_L$ & U(1)$_Y$ & SU(2)$_H$ & U(1)$_R$ \\
\hline 
 $Q$        & 3         & 2         & 1/6      & 1         & 1        \\
 $\bar U$   & 3$^*$     & 1         & -2/3     & 1         & 1        \\
 $\bar D$   & 3$^*$     & 1         & 1/3      & 1         & 1        \\
 $L$        & 1         & 2         & -1/2     & 1         & 1        \\
 $\bar N$   & 1         & 1         & 0        & 1         & 1        \\
 $\bar E$   & 1         & 1         & 1        & 1         & 1        \\
 $H$        & 1         & 2         & 1/2      & 1         & 0        \\
 $\bar H$   & 1         & 2         & -1/2     & 1         & 0        \\
\hline
 $H'$       & 1         & 2         & 1/2      & 1         & 2        \\
 $\bar H'$  & 1         & 2         & -1/2     & 1         & 2        \\
 $\Omega_i$ & 8         & 1         & 0        & 1         & 3/4      \\
 $\Sigma_i$ & 1         & 3         & 0        & 1         & 3/4      \\
 $X$        & 1         & 1         & 1        & 1         & 1/2      \\
 $X'$       & 1         & 1         & 1        & 1         & 3/2      \\
\hline
 $Q_1$      & 1         & 1         & 1        & 2         & $q_1$    \\
 $Q_2$      & 1         & 1         & 1        & 2         & $q_2$    \\
 $S$        & 1         & 1         & 1        & 1         & 4        \\
\end{tabular}
\caption{
Particle contents of the model.
The system of the fields $Q_1$, $Q_2$ and $S$
 with SU(2)$_H$ gauge symmetry constitutes
 the hidden (sequestered) sector.
Other fields are the member of the visible sector.
The index $i$ runs from 1 to 4
 for $\Omega_i$ and $\Sigma_i$.
The charges $q_1$ and $q_2$
 follow the constraint of $q_1 + q_2 = -2$.
The concrete values of these charges
 are determined through the cancellations of
 $(\mbox{\rm U(1)}_R)^3$ and $\mbox{\rm U(1)}_R (\mbox{\rm gravity})^2$
 anomalies (see text).
}
\label{contents}
\end{table}


\begin{references}
\bibitem{polonyi}
 J.~Polonyi, Budapest preprint KFKI-93 (1977).
\bibitem{randall-sundrum1}
 L.~Randall and R.~Sundrum, Nucl. Phys. B557, 79 (1999).
\bibitem{GLMR}
 G.F.~Giudice, M.A.~Luty, H.~Murayama and R.~Rattazzi,
 JHEP 9812, 027 (1998).
\bibitem{BMP}
 J.A.~Bagger, T.~Moroi and E.~Poppitz, JHEP 0004, 009 (2000).
\bibitem{pomarol-rattazzi}
 A.~Pomarol and R.~Rattazzi, JHEP 9905, 013 (1999).
\bibitem{CLMP}
 Z.~Chacko, M.A.~Luty, I.~Maksymyk and E~Ponton,
 JHEP 0004, 001 (2000). 
\bibitem{KSS}
 E.~Katz, Y.~Shadmi and Y.~Shirman, JHEP 9908, 015 (1999).
\bibitem{chamseddine-dreiner}
 A.H.~Chamseddine and H.~Dreiner, Nucl. Phys. B458, 65 (1996).
\bibitem{CFM}
 D.J.~Casta\~no, D.Z.~Freedman and C.~Manuel,
 Nucl. Phys. B461, 50 (1996).
\bibitem{JJ}
 I.~Jack and D.R.T.~Jones, Phys. Lett. B482, 167 (2000).
\bibitem{KMO}
 N.~Kitazawa, N.~Maru and N.~Okada,
 hep-ph/9911251 to be published in Phys. Rev. D;
 hep-ph/0003240 to be published in Nucl. Phys. B.
\bibitem{IKYY}
 K.~Inoue, M.~Kawasaki, M.~Yamaguchi and T.~Yanagida,
 Phys. Rev. D45, 328 (1992).
\bibitem{kaku-townsend}
 M.~Kaku and P.K.~Townsend, Phys. Lett. 76B, 54 (1978)
\bibitem{kugo-uehara}
 T.~Kugo and S.~Uehara,
 Nucl. Phys. B222, 125 (1983); {\it ibid} B226, 49 (1983).
\bibitem{FGKP}
 S.~Ferrara, L.~Girardello, T.~Kugo and A.~Van Proeyen,
 Nucl. Phys. B223, 191 (1983).
\bibitem{CFGP}
 E.~Cremmer, S.~Ferrara, L.~Girardello and A.~Van Proeyen,
 Nucl. Phys. B212, 413 (1983).
\bibitem{radiative}
 L.~Iba\~nez and G.G.~Ross, Phys. Lett. 110B, 215 (1982);
 K.~Inoue, A.~Kakuto, H.~Komatsu and S.~Takeshita,
 Prog. Theor. Phys. 68, 927 (1982);
 L.~Alvarez-Gaume, M.~Claudson and M.B.~Wise,
 Nucl. Phys. B207, 96 (1982).
\bibitem{ADS}
 I.~Affleck, M.~Dine and N.~Seiberg, Nucl. Phys. B241, 493 (1984).
\bibitem{extra-U(1)}
 S.~Weinberg, Phys. Rev. D26, 287 (1982);
 L.J.~Hall and I.~Hinchliffe, Phys. Lett. 112B, 351 (1982).
\bibitem{MP}
 E.A.~Mirabelli and M.E.~Peskin, Phys. Rev. D58, 065002 (1998).
\end{references}
\end{document}